\documentclass[twoside,11pt]{article}

\usepackage[preprint]{jmlr2e}
\hypersetup{colorlinks=true,linkcolor=blue,citecolor=blue,urlcolor=blue}
\usepackage{amsmath,mathtools,relsize}
\usepackage{mathrsfs,stmaryrd}
\usepackage{bm}
\usepackage{bbm}
\usepackage{enumerate}
\usepackage{enumitem}
\usepackage{rotating}
\usepackage{float}
\usepackage{booktabs}
\usepackage[ruled,noend]{algorithm2e}
\usepackage{epstopdf}
\usepackage{lastpage}

\newtheorem{lemma}{Lemma}

\newtheorem{proposition}{Proposition}

\newtheorem{remark}{Remark}

\newtheorem{corollary}{Corollary}

\newtheorem{definition}{Definition}
\newtheorem{condition}{Condition}

\jmlrheading{0}{2026}{1-\pageref{LastPage}}{TBD}{TBD}{TL-HFS}{Zejing Zheng, Rui Huang and Junlong Zhao}
\ShortHeadings{Transfer Learning with Heterogeneous Feature Spaces}{Zheng, Huang and Zhao}
\firstpageno{1}

\newcommand{\bB}{{\boldsymbol B}}

\newcommand{\bG}{{\boldsymbol G}}

\newcommand{\bI}{{\boldsymbol I}}

\newcommand{\bv}{{\boldsymbol v}}

\newcommand{\bx}{{\boldsymbol x}}
\newcommand{\bX}{{\boldsymbol X}}

\newcommand{\bz}{{\boldsymbol z}}

\newcommand{\bdelta}{{\boldsymbol \delta}}
\newcommand{\bbeta}{{\boldsymbol \beta}}

\newcommand{\blambda}{{\boldsymbol \lambda}}
\newcommand{\btheta}{{\boldsymbol \theta}}

\newcommand{\btau}{{\boldsymbol \tau}}

\newcommand{\bSig}{{\boldsymbol \Sigma}}

\newcommand{\bGamma}{{\boldsymbol \Gamma}}

\newcommand{\mE}{\mathbb{E}}
\newcommand{\mS}{\mathcal{S}}

\newcommand{\mR}{\mathbb{R}}

\newcommand{\diag}{\mathrm{diag}}

\newcommand{\argmin}{\mathrm{argmin}}

\newcommand{\mT}{{\mathcal{T}}}
\newcommand{\tbX}{{\widetilde \bX}}
\newcommand{\tbx}{{\tilde \bx}}
\newcommand{\mI}{{\mathcal{I}}}
\newcommand{\bdot}{\mathsmaller{\bullet}}
\newcommand{\hiw}{{\rm hiw}}
\makeatletter

\newcommand{\Rmnum}[1]{\expandafter\@slowromancap\romannumeral #1@}
\makeatother

\begin{document}

\title{Transfer Learning with Heterogeneous Feature Spaces in Linear Regression}

\author{\name Zejing Zheng \email zzjstat23@mail.bnu.edu.cn \\
       \addr School of Statistics\\
       Beijing Normal University\\
       19 Xinjiekouwai Street, Haidian District\\
       Beijing 100875, China
       \AND
       \name Rui Huang \email 202421011017@mail.bnu.edu.cn \\
       \addr School of Statistics\\
       Beijing Normal University\\
       19 Xinjiekouwai Street, Haidian District\\
       Beijing 100875, China
       \AND
       \name Junlong Zhao\textsuperscript{*} \email zhaojunlong928@126.com \\
       \addr School of Statistics\\
       Beijing Normal University\\
       19 Xinjiekouwai Street, Haidian District\\
       Beijing 100875, China}

\editor{TBD}

\maketitle
\begingroup
\renewcommand{\thefootnote}{\fnsymbol{footnote}}
\footnotetext[1]{Corresponding author.}
\endgroup

\begin{abstract}
Transfer learning improves target-task performance by leveraging related source data. Most methods assume shared feature spaces, yet in many applications, each source observes only a subset of target covariates. Classical imputation fails here due to block missingness, and standard imputation matrices are not optimized for target parameter estimation. We study low- and high-dimensional linear regression and propose Heterogeneous Importance Weighting (HIW). Our method aligns feature spaces via projection-based imputation and transfers information through sample-selected importance weighting. This framework accommodates diverse projection matrices to construct target-oriented imputation. We develop a classification-based procedure with pseudo-responses to estimate conditional error densities for the weights. We establish entry-wise and global convergence rates for the estimator, with numerical and real-data studies demonstrating its effectiveness.
\end{abstract}


\section{Introduction}
Transfer learning is an important tool in statistics and machine learning to improve the performance of a target task by borrowing information from related source data, especially when the target sample size is limited \citep{weiss2016survey}.
It has been successfully applied in various fields, such as computer vision and medical diagnosis \citep{daume2009frustratingly,hajiramezanali2018bayesian}.
Recent statistical developments have studied transfer learning in a variety of models, including linear regression, generalized linear models, and many others \citep[etc.]{bastani2021predicting, li2022transfer, hu2023optimal, tian2023transfer, li2024estimation, he2024adatrans, ren2024transfer, yang2025precise, zhao2026residual, cai2021transfer, li2023transfer, cai2024transfer}.
Most of these, however, are developed under the assumption that the target and sources share the same feature space.

In practice, the homogeneous feature space assumption is often violated. 
For a regression problem, let $\{X_j:j\in\Theta_0\}$ denote the covariates observed in the target task and $\{X_j:j\in\Theta_k\}$ those observed in source $k$, where $\Theta_0,\Theta_k\subseteq\{1,\ldots,p\}$ for $k=1,\ldots,K$. 
Heterogeneous feature spaces arise when $\Theta_k\neq\Theta_0$ for at least one source. 
Such heterogeneity is common in applications: image data collected at different resolutions can yield different feature representations \citep{duan2012learning}; 
electronic health records from various healthcare centers often contain different clinical variables because of differences in data systems and documentation practices \citep{wang2022survmaximin}.

In this paper, we study transfer learning for linear regression models with heterogeneous feature spaces (HFS).
We focus on the setting where the target contains the full feature set, $\Theta_0=\{1,\ldots,p\}$, while the $k$-th source observes a subset of the target features, $\Theta_k\subseteq\Theta_0$ for $k=1,\ldots,K$.
This formulation allows different sources to contain different subsets of target features; sources sharing the full target feature space can also be handled.

This HFS setting poses several challenges for transfer learning.  
 First, classical imputation methods are not directly suited to the HFS setting.
One difficulty is structural: some source covariates are entirely unobserved, leading to a block-missing pattern for which standard imputation methods are difficult to apply.
Another difficulty is objective mismatch.
Classical imputation is typically source-oriented, in the sense that it aims to recover missing covariates or improve prediction within the same data source \citep{little2019statistical,stekhoven2012missforest}.
In contrast, in transfer learning, the imputation should be target-oriented rather than source-oriented.
Moreover, the imputation may introduce additional bias and lead to negative transfer, in which source-domain information worsens rather than improves performance on the target domain \citep{weiss2016survey}.

Second, standard transfer learning methods designed for homogeneous feature spaces cannot be directly adopted in the HFS setting. 
For homogeneous linear models, importance weights can be constructed using a one-dimensional error density ratio \citep{zhao2026residual}.
In contrast, under HFS, weight construction involves conditional density ratios that may be multivariate or high-dimensional, which makes both the formulation and estimation of the weights substantially more challenging.

Third, the theoretical analysis  becomes more complex.
Since the observed feature sets vary across sources, some covariates may appear in many sources, whereas others may appear in only a few.
Thus, different covariates commonly have different sample sizes.
It is therefore desirable to establish entry-wise convergence rates in addition to global error bounds, which have received relatively limited attention in the existing literature.

There are several studies on transfer learning in linear models under HFS.
\citet{zhao2023heterogeneous} explored scenarios with a single source and overlapping feature spaces (i.e., $\Theta_0 \cap \Theta_1\neq\emptyset$), where only the summary statistics from the source are available.
\citet{chang2024heterogeneous} studied the case $\Theta_0 \subsetneqq \Theta_k$, where the source contains features unavailable in the target.
This method imputes missing target features via a feature map learned from source data, which requires the source and target feature maps to be sufficiently similar. However, this condition is difficult to verify and often does not hold in practical applications.
Moreover, both approaches lack a process for selecting sources or samples, potentially leading to negative transfer.
Finally, neither method studies entry-wise convergence rates.

\subsection{Contributions}
Motivated by these gaps, we propose Heterogeneous Importance Weighting (HIW) for linear models under HFS, covering both low- and high-dimensional settings.
The main contributions of this work are summarized as follows.

First, we propose a framework based on importance weighting that is theoretically valid for a broad class of projection matrices $\bG^{(k)}$ at the population level (see Proposition~\ref{prop: iwtl} and Remark~\ref{rem: iwtl}). 
The flexibility of $\bG^{(k)}$ enables us to construct target-oriented imputation matrices and further boost target-task performance.  
To achieve this goal, we treat projection matrices as tunable hyperparameters and build a data-fused projection matrix that aggregates information from source and target domains alike, followed by a refinement step to yield its target-specific variant. 
Furthermore, combining importance weighting with sample selection enables stable and effective knowledge transfer and helps mitigate the risk of negative transfer.

Second, computing the weight requires estimating a conditional density, which is often challenging in practice. 
To handle this, we propose a classification-based framework that avoids direct estimation of the conditional error density. 
In conventional classifier-based density-ratio methods \citep{bickel2007discriminative,sugiyama2012density}, samples from both the numerator and denominator distributions are observed. 
Our setting is fundamentally different: there are no real observations from the numerator distribution.
We address this issue by generating pseudo-responses from the specified reference distribution, thereby constructing samples corresponding to the numerator distribution.

Third, we establish convergence rates for the proposed estimator at both the global and entry-wise levels.
Let $\bbeta^{(0)}=(\beta_1^{(0)},\ldots,\beta_p^{(0)})^\top$ denote the true target coefficient vector, and let $\widehat\bbeta^{(0)}=(\hat\beta_1^{(0)},\ldots,\hat\beta_p^{(0)})^\top$ denote its estimator. 
For concreteness, under the low-dimensional regime, the squared entry-wise estimation error $(\hat\beta_j^{(0)}-\beta_j^{(0)})^2$ scales proportionally to $1/N_j$, where the effective sample size is defined as
$N_j=n_0+N_j^{\text{dir}}+N_j^{\text{imp}}$. 
In this expression, $n_0$ refers to the sample size of the target dataset, $N_j^{\text{dir}}$ denotes the effective sample size contributed by sources in which $X_j$ is directly observed, and $N_j^{\text{imp}}$ corresponds to the effective sample size from sources where $X_j$ is imputed. 
From this result, the associated global error rate scales proportionally to $\sum_{j=1}^p(1/N_j)$. 
Analogous convergence rates hold in high-dimensional regimes, up to logarithmic factors.

\subsection{Organization and notation}
The remainder of this paper is structured as follows.
Section \ref{sec: transfer} introduces our proposed HIW method within the HFS framework.
Section \ref{sec: theory HIW} derives the theoretical guarantees of the method for both low- and high-dimensional regimes.
Sections \ref{sec: sim} and \ref{sec: el} report simulation experiments and real-data applications, respectively.
Section \ref{sec: dis} concludes the paper.

We summarize the notation used throughout the paper.
Let $a\wedge b=\min\{a,b\}$ and, for an integer $K>0$, let $[K]=\{1,\ldots,K\}$.
For two positive sequences $\{a_n\}$ and $\{b_n\}$, write $a_n\gtrsim b_n$ or $a_n=\Omega(b_n)$ if $\liminf_{n\to\infty}a_n/b_n>0$, write $a_n\lesssim b_n$ or $a_n=O(b_n)$ if $b_n\gtrsim a_n$, and write $a_n\asymp b_n$ if both hold; $a_n\ll b_n$ and $a_n=o(b_n)$ both mean $a_n/b_n\to0$.
The stochastic notations $o_p(\cdot)$ and $O_p(\cdot)$ are defined analogously.
For a vector $\bv$, $\|\bv\|_0$, $\|\bv\|_1$, $\|\bv\|_2$, and $\|\bv\|_\infty$ denote the usual $\ell_0$, $\ell_1$, $\ell_2$, and $\ell_\infty$ norms.
For a matrix $\bB$, $\interleave\bB\interleave_1$ and $\interleave\bB\interleave_2$ denote its $1$-norm and $2$-norm, and $\bB_{A_1,A_2}$, $\bB_{\bdot,A_2}$, and $\bB_{A_1,\bdot}$ denote the corresponding submatrix, columns, and rows indexed by $A_1$ and $A_2$.
For two sets $A_1$ and $A_2$, $A_1\subseteq A_2$ denotes inclusion and $A_1\subsetneqq A_2$ denotes strict inclusion.
\section{Transfer learning under heterogeneous feature spaces}\label{sec: transfer}
This section covers both low- and high-dimensional settings. In the low-dimensional regime, the covariate dimension is fixed or much smaller than the sample size; in the high-dimensional regime, the covariate dimension is much larger than the sample size.
\subsection{Heterogeneous feature spaces and data imputation}\label{subsec: imputation}
Heterogeneous feature spaces can be viewed as arising from partially observed covariates in an underlying homogeneous feature space.
For exposition, we first introduce the full linear models under a common feature set, with $\Theta_k=[p]$ for $k=0,1,\ldots,K$.
The target data $(k=0)$ and source data $(k\in[K])$ are assumed to be independently generated from
\begin{equation}\label{eq:target model}
   \setlength\abovedisplayskip{10pt}
 \setlength\belowdisplayskip{10pt}
  {y}_{i}^{(k)}=(\bx_{i}^{(k)})^{\top}\bbeta^{(k)}+{\epsilon}_{i}^{(k)}, ~\; i=1,\ldots,n_k, ~\; k=0,1,\ldots,K.
\end{equation}
Here, $\bx_{i}^{(k)}=(x_{i1}^{(k)},\ldots,x_{ip}^{(k)})^{\top} \in \mathbb{R}^{p}$ are i.i.d. random vectors with mean zero and covariance matrix $ \bSig^{(k)}=(\sigma^{(k)}_{j,l}) \in \mathbb{R}^{p\times p}$, and $y_{i}^{(k)} \in \mathbb{R}$ is the response variable.
The regression coefficient vector is denoted by $\bbeta^{(k)} = (\beta_{1}^{(k)}, \ldots, \beta_{p}^{(k)})^{\top} \in \mathbb{R}^p$, and the random errors $\epsilon_{i}^{(k)}$ are i.i.d. with mean zero and are independent of $\bx_{i}^{(k)}$.

We next describe the observed data under HFS. Recall that we focus on the case  $\Theta_k \subseteq \Theta_0=[p]$ for $1\le k\le K$;   the cardinality of $\Theta_k$ is denoted by $p_k = |\Theta_k|$. 
 We refer to the setting as strong HFS if $\Theta_k\subsetneqq\Theta_0$ for all $k\in[K]$, and as weak HFS if at least one source satisfies $\Theta_k=\Theta_0$.
We focus on strong HFS in the main development; weak HFS can be handled with only minor modifications, as discussed in Remark~\ref{remark: weak HFS}.

Under strong HFS, we observe only the subvector $\tbx_{i}^{(k)}=(x^{(k)}_{ij})_{j\in \Theta_k}^\top \in\mR^{p_k}$ instead of the full vector $\bx_i^{(k)}$ for $i\in [n_k]$.  
The corresponding reduced linear model is
\begin{equation}\label{eq:source model pl}
  {y}_{i}^{(k)} = (\tbx_{i}^{(k)})^{\top}{\btheta}^{(k)} + \tilde{\epsilon}_{i}^{(k)}, \quad i=1,\ldots,n_k, \quad k\in[K],
\end{equation}
where ${\btheta}^{(k)} \in \mathbb{R}^{p_k}$ is the marginal coefficient vector, and $\tilde{\epsilon}_{i}^{(k)}$ is the reduced-model error with mean zero.
In general, $\tilde{\epsilon}_{i}^{(k)}$ differs from the error $\epsilon_i^{(k)}$ in \eqref{eq:target model} and may depend on the observed covariates $\tbx_i^{(k)}$. 
Let $\tbx^{(k)}$ and $\tilde{\epsilon}^{(k)}$ denote the population counterparts of $\tbx_i^{(k)}$ and $\tilde{\epsilon}_i^{(k)}$, respectively. 
Likewise, let $\bx^{(k)}$, $y^{(k)}$, and $\epsilon^{(k)}$ denote the population counterparts of $\bx_i^{(k)}$, $y_i^{(k)}$, and $\epsilon_i^{(k)}$, respectively.
Then we impose the following regularity condition.
\begin{condition}\label{Con: covariance epsilon}
  \begin{itemize}
\item[(a)] The vectors $\bx^{(0)}$ and $\tbx^{(k)}$ are sub-Gaussian random vectors, and the errors $\epsilon^{(0)}$ and $\tilde{\epsilon}^{(k)}$ are sub-Gaussian random variables, for all $k\in[K]$.
\item[(b)] The smallest and largest eigenvalues of $\bSig^{(0)}$ and $\bSig_{\Theta_k,\Theta_k}^{(k)}$ are bounded away from zero and infinity for all $k\in[K]$.
\end{itemize}
\end{condition}

Condition \ref{Con: covariance epsilon} is a variant of the standard regularity condition used in the full-covariate setting \citep[see, e.g.,][]{li2022transfer, he2024adatrans}. The main difference is that, in the HFS setting, the condition for the source domain is imposed on the observed covariates under the reduced model rather than on the full covariate vector.

To estimate the target parameter $\bbeta^{(0)}$ using both the target data and the observed source data, we first align the covariate dimensions by imputing the unobserved source covariates.
Specifically, given any projection matrix $\bG^{(k)} \in \mathbb{R}^{p_k \times p}$, the imputed source covariate vector is defined by 
$$
   \setlength\abovedisplayskip{8pt}
 \setlength\belowdisplayskip{8pt}
\bx_{\mathrm{im}}^{(k)} =(\bG^{(k)})^{\top} \tbx^{(k)}\in \mathbb{R}^p.
$$
Throughout the paper, we assume that the columns of the projection matrices are nondegenerate, in the sense that $\|\bG_{\bdot,j}^{(k)}\|_2$ is bounded away from zero for all $k\in[K]$ and $j\in[p]$.

Standard imputation methods, which are not developed in the context of transfer learning, generally seek to predict missing values from observed data within the same dataset \citep{little2019statistical,stekhoven2012missforest}.
For example, at the population level, a common choice is  $\bG^{(k)}=(\bSig^{(k)}_{\Theta_k,\Theta_k})^{-1}\bSig^{(k)}_{\Theta_k,\bdot}$, the linear projection coefficient of the full source covariate vector on the observed source covariates.

This construction, however, is not adequate for our transfer learning objective. 
First, because covariates outside $\Theta_k$ are not observed in source $k$, the cross-covariance matrix $\bSig^{(k)}_{\Theta_k,\bdot}$ cannot be estimated from source $k$ alone. 
Second, even if this population matrix were known, a projection optimized for predicting the unobserved source covariates may not be optimal for estimating the target parameter $\bbeta^{(0)}$.
We therefore treat $\bG^{(k)}$ as a tuning parameter to be chosen for transfer, with its practical construction presented  in Section~\ref{sec: projection matrix}.

 \begin{remark}\label{remark: weak HFS}
Under weak HFS, sources sharing the full target feature space can be handled.
For any source $k'$ with $\Theta_{k'}=\Theta_0$, no imputation is needed, and we simply set $\bG^{(k')}=\bI_p$, where $\bI_p$ is the $p\times p$ identity matrix.
\end{remark}

In general, it is impossible to completely eliminate the heterogeneity between the imputed source covariates and the target covariates, regardless of the choice of the projection matrix. Consequently, incorporating source data through imputation may still introduce two sources of bias: imputation bias, arising from replacing unobserved covariates with imputed values, and transfer bias, resulting from discrepancies between the source and target regression coefficients. These biases may lead to severe negative transfer if left unaddressed. Importance weighting has been shown to be an effective strategy for mitigating distributional discrepancies in transfer learning \citep[etc.]{zhao2026residual,zheng2025transfer}. Under the HFS setting, however, constructing appropriate importance weights is considerably more challenging, as discussed in the next subsection.

\subsection{Importance weighting framework under heterogeneous feature spaces}\label{subsec: IWTL}
We first present the importance weighting framework under a simple setting $[p]=\cup_{k=1}^K\Theta_k$, that is, the union of the observed source feature sets equals the target feature space. This setting can be relaxed, as discussed in Remark~\ref{rem: iwtl}.
Let $f_k(\tilde{\epsilon}^{(k)}\mid \tbx^{(k)})$ denote the conditional density of the reduced-model error $\tilde{\epsilon}^{(k)}$ given the observed source covariates $\tbx^{(k)}$ for $k\in[K]$.
Let $\nu$ be a researcher-specified {\it reference error} with density $f_\nu$, satisfying part (b) of Condition~\ref{Con: f epsilon} below, and  define  the ideal response $y^{(k0)}=(\tbx^{(k)})^{\top}\bG^{(k)}\bbeta^{(0)}+\nu$ under the parameter $\bbeta^{(0)}$ \citep{zhao2026residual}. 
To transfer the conditional distributions of $y^{(k)}|\tbx^{(k)}$ to $y^{(k0)}|\tbx^{(k)}$, we define the 
   HFS importance weight for source $k$ as
\begin{equation*}
\omega^{(k)} =\frac{f_\nu (y^{(k)}-(\tbx^{(k)})^{\top}\bG^{(k)}\bbeta^{(0)})}{f_k( \tilde{\epsilon}^{(k)} \mid \tbx^{(k)})}= \frac{f_\nu (y^{(k)}-(\tbx^{(k)})^{\top}\bG^{(k)}\bbeta^{(0)})}{f_k(y^{(k)}-(\tbx^{(k)})^{\top}{\btheta}^{(k)} \mid \tbx^{(k)})}.
\end{equation*}
In practice, the unknown parameters involved (e.g. $\bbeta^{(0)},\btheta^{(k)}$)  can be replaced by their initial estimators. We provide the following discussion on our weighting scheme.

First, this specifically designed weight significantly mitigates the risk of negative transfer. Our method ensures validity under only mild conditions on the projection matrix  $\bG^{(k)}$ (as demonstrated in Proposition \ref{prop: iwtl} below). Furthermore, this flexibility in choosing  $\bG^{(k)}$ allows us to construct a target-oriented projection matrix, thereby enhancing the estimation efficiency for the target parameter (see Section \ref{sec: projection matrix}).

Second, the denominator of our weight is a conditional density function, which is challenging to estimate, particularly in high-dimensional settings. 
This difficulty arises because the reduced-model error $\tilde\epsilon^{(k)}$ generally depends on the observed source covariates $\tbx^{(k)}$. 
To avoid direct conditional-density estimation, we introduce in Section~\ref{subsec: iwtl_practice} a classification-based approach based on constructed pseudo-responses. 
This weight also requires a separate construction and analysis of the projection matrix $\bG^{(k)}$.

Third, standard importance weights are typically defined as ratios of the joint densities of $(y^{(0)},(\bx^{(0)})^\top)$ and $(y^{(k)},(\bx^{(k)})^\top)$. Such ratios are not directly available under HFS because some source covariates are unobserved. Even if all covariates were observed, estimating these joint density ratios would remain difficult in high-dimensional settings.

For clarity, we impose the following regularity conditions on the density functions.
\begin{condition}\label{Con: f epsilon}
\begin{itemize}
    \item[(a)] For each $k\in[K]$, the conditional density function $f_k(t\mid \tbx^{(k)})$ is upper bounded and positive for all finite $t$.
    \item[(b)] The random variable $\nu$ is independent of the source data and satisfies $\mE(\nu)=0$. Its density function $f_\nu(t)$ is symmetric, bounded, and positive for all finite $t$.
\end{itemize}
\end{condition}

Condition \ref{Con: f epsilon} is mild and is commonly satisfied in practice. 
In particular, part (a) is satisfied when the conditional error $\tilde{\epsilon}^{(k)}|\tbx^{(k)}$ follows many commonly encountered distributions (e.g., normal or $t$ distributions). Part (b) is mild, as $\nu$ is specified by the user. For instance, one can choose $\nu$ to be standard normal or a centered $t$ distribution.

Let $\bB\in\mR^{p\times p}$ be a positive-definite working matrix and define $P_\bB^{(k)}=(\bB_{\Theta_k,\Theta_k})^{-1}\bB_{\Theta_k,\bdot}$.
In Proposition~\ref{prop: iwtl} below, we take $\bG^{(k)}=P_\bB^{(k)}$ for $k\in[K]$; this choice is made for clarity and can be relaxed, as discussed in Remark~\ref{rem: iwtl}. 
For each source $k\in[K]$, let $\tilde{\bz}^{(k)}=(y^{(k)},(\tbx^{(k)})^\top)$.
For the target domain, set $\omega^{(0)}=1$, $\tbx^{(0)}=\bx^{(0)}$, $\tilde{\bz}^{(0)}=(y^{(0)},(\bx^{(0)})^\top)$, and $\bG^{(0)}=\bI_p$.
Let $N=\sum_{k=0}^K n_k$ and $\alpha_k=n_k/N>0$ for $k=0,1,\ldots,K$.
We next provide a population-level justification for the proposed importance weighting framework.
\begin{proposition}\label{prop: iwtl}
 Assume that the weights $\omega^{(k)}$ are bounded away from zero for $k=1,\ldots,K$.  
  Let $\bG^{(k)}=P_\bB^{(k)}$ for $k=1,\ldots,K$. Under Conditions \ref{Con: covariance epsilon} and \ref{Con: f epsilon}, the target parameter $\bbeta^{(0)}$ satisfies
\begin{equation*}
    \bbeta^{(0)} = \underset{\bbeta \in \mR^{p}}{\argmin} \sum_{k=0}^{K} \alpha_k \mathbb{E}_{\tilde{\bz}^{(k)}} \left[ \omega^{(k)}(y^{(k)}-(\tbx^{(k)})^{\top}\bG^{(k)}\bbeta)^2 \right].
\end{equation*}
\end{proposition}

\begin{remark}\label{rem: iwtl} 
The condition $[p]=\cup_{k=1}^K\Theta_k$ and the explicit  form of $\bG^{(k)}$ in Proposition~\ref{prop: iwtl} can be relaxed.
In fact, it suffices to assume that the minimum eigenvalue of $\bGamma$ is bounded away from zero, where
\begin{equation}\label{eq: gamma}
	\setlength\abovedisplayskip{8pt}
	\setlength\belowdisplayskip{8pt}
	\bGamma = \alpha_0\mE[\bx^{(0)}(\bx^{(0)})^\top]+\sum_{k=1}^K \alpha_k \mE[\bx_{\mathrm{im}}^{(k)}(\bx_{\mathrm{im}}^{(k)})^\top],
\end{equation}
is the covariance matrix of the mixture distribution of the target covariates $\bx^{(0)}$ and the imputed source covariates $\bx_{\mathrm{im}}^{(k)}$.
Since each $\bx_{\mathrm{im}}^{(k)}$ depends on $\bG^{(k)}$, this condition implicitly imposes constraints on the projection matrices. See the proof of Proposition~\ref{prop: iwtl} in Appendix~A for details.
\end{remark}

Proposition~\ref{prop: iwtl} shows that the proposed importance-weighting scheme can recover $\bbeta^{(0)}$ by combining the target data with the imputed source data.
This population-level validity holds for any positive-definite working matrix $\bB$, indicating that the method is robust to the choice of imputation matrix. 
In finite samples, however, the statistical efficiency depends on $\bB$, which motivates the data-driven construction of $\bB$ in Section~\ref{sec: projection matrix}.

For stability, we use only source observations on which the weights are uniformly bounded.
Let $\eta_i^{(k)} = (\tbx_i^{(k)})^{\top}(\boldsymbol{\theta}^{(k)}-\bG^{(k)}\bbeta^{(0)})$ for $i \in [n_k]$ and $k \in [K]$. 
Define the sample-level importance weight and the corresponding selection set $\mathcal{I}_k$ by
$$
\begin{aligned}
  &\omega_i^{(k)}={f_\nu (y_i^{(k)}-(\tbx_i^{(k)})^{\top}\bG^{(k)}\bbeta^{(0)})}/{f_{k}(y_i^{(k)}-(\tbx_i^{(k)})^{\top}{\btheta}^{(k)} \mid \tbx_i^{(k)})},\\
&\mI_k = \{i \in [n_k] : |y_i^{(k)}-(\tbx_i^{(k)})^{\top}\bG^{(k)}\bbeta^{(0)}| \leq R,\quad |\eta_i^{(k)}| \leq M_k\},
\end{aligned}
$$
where $R>0$ and $M_k>0$ are tuning parameters. 
Only observations in $\mathcal I_k$ are transferred from source $k$.
By Lemma~A.2 in Appendix~A, if $R$ and $M_k$ are finite, then the weights $\omega_i^{(k)}$ are bounded away from zero and infinity for all $i\in\mathcal I_k$. 
In implementation, $R$ and $M_k$ are selected by cross-validation, with $R$ typically chosen to be large.
The next subsection describes how the unknown weights and selected sets are estimated.

\subsection{Importance weighting transfer learning in practice} \label{subsec: iwtl_practice}
In this subsection, we first estimate the weights $\omega^{(k)}$ with a given $\bG^{(k)}$ and the selection sets $\mathcal{I}_k$ for $k \in [K]$, and then present the practical implementation of the importance weighting transfer learning method under HFS.
As discussed in Section \ref{subsec: IWTL}, directly estimating the conditional density $f_k(\tilde{\epsilon}^{(k)} \mid \tbx^{(k)})$ in the weight $\omega^{(k)}$ is difficult, especially in high-dimensional settings.

Let $\bm{\xi}$ and $\tilde{\bm{\xi}}$ be two $p$-dimensional random vectors with density functions $f_{\bm{\xi}}$ and $f_{\tilde{\bm{\xi}}}$, respectively. When $p$ is moderate or large, the estimation of the density ratio $f_{\bm{\xi}} / f_{\tilde{\bm{\xi}}}$ is often framed as a binary classification problem to distinguish samples from the two distributions \citep{bickel2007discriminative, sugiyama2012density}. However, this standard framework cannot be directly applied here, as the denominator of $\omega^{(k)}$ involves a conditional error density, whereas the numerator is defined by a predefined reference distribution for which no corresponding samples are observed. To bridge this gap, we generate pseudo-responses $y^{(k0)} = (\tbx^{(k)})^\top\bG^{(k)}\bbeta^{(0)} + \nu$, where $\nu$ is the reference noise. By training a classifier to discriminate between the synthetic observations $(y^{(k0)}, \tbx^{(k)})$ and the observed data $(y^{(k)}, \tbx^{(k)})$, we recover the weight via the classifier's odds. This approach is formally established in Proposition~\ref{prop: weight_optimality}.

\begin{proposition}\label{prop: weight_optimality}
For each source $k \in [K]$, let $P_k$ denote the joint distribution of $\tilde{\bz}^{(k)} = (y^{(k)}, (\tbx^{(k)})^{\top})$, and let $P_{k0}$ denote the synthetic distribution of $(y^{(k0)}, (\tbx^{(k)})^{\top})$, where $y^{(k0)} = (\tbx^{(k)})^{\top}\bG^{(k)}\bbeta^{(0)} + \nu$ with the random variable $\nu$ satisfying part (b) of Condition~\ref{Con: f epsilon}.
For any probabilistic classifier $D(\tilde{\bz}) \in (0,1)$ where $\tilde{\bz} \in \mR^{p_k+1}$, define the expected cross-entropy risk as
$$
     \setlength\abovedisplayskip{8pt}
 \setlength\belowdisplayskip{8pt}
  \mathcal{R}(D) = - \mathbb{E}_{\tilde{\bz} \sim P_{k0}} \big[\log D(\tilde{\bz})\big] - \mathbb{E}_{\tilde{\bz} \sim P_k} \big[\log (1 - D(\tilde{\bz}))\big].
$$
Then the minimizer $D^{(k)}(\tilde{\bz}) = \arg\min_D \mathcal{R}(D)$ satisfies
\begin{equation}\label{eq: odds_ratio}
    \frac{D^{(k)}(\tilde{\bz}^{(k)})}{1 - D^{(k)}(\tilde{\bz}^{(k)})} = \frac{f_\nu \big(y^{(k)}-(\tbx^{(k)})^{\top}\bG^{(k)}\bbeta^{(0)}\big)}{f_k\big(y^{(k)}-(\tbx^{(k)})^{\top}{\btheta}^{(k)}|\tbx^{(k)}\big)} = \omega^{(k)}.
\end{equation}
\end{proposition}

Proposition~\ref{prop: weight_optimality} shows that the desired weight can be recovered from the odds of the optimal classifier. 
In practice, this classifier can be estimated by a binary classification problem in which samples from $P_{k0}$ are assigned label $L=1$ and samples from $P_k$ are assigned label $L=0$. 
We next use this characterization to construct a practical three-step estimation procedure.

{\it Step 1}. Construct initial estimators $\bbeta_{\rm in}^{(0)}$ and $\btheta_{\rm in}^{(k)}$ for $\bbeta^{(0)}$ and $\btheta^{(k)}$, respectively, using the target and source datasets. The estimation approach is tailored to the dimensionality: in low-dimensional settings, ordinary least squares (OLS) can be employed, whereas in high-dimensional settings, regularized estimators such as the LASSO \citep{tibshirani1996regression} are preferred.

{\it Step 2}. Generate a synthetic sample based on Proposition \ref{prop: weight_optimality}. More precisely, conditional on the observed covariates $\tbx^{(k)}$, we generate pseudo-responses according to
$$
    \setlength\abovedisplayskip{8pt}
 \setlength\belowdisplayskip{8pt}
y^{(k0)}=(\tbx^{(k)})^\top \bG^{(k)}\bbeta_{\rm in}^{(0)}+\nu,
$$
where $\nu$ satisfies part (b) of Condition~\ref{Con: f epsilon}; for example, $\nu$ can be taken to follow a standard normal distribution.
Then the resulting pairs $(y^{(k0)},(\tbx^{(k)})^{\top})$ are labeled as $L=1$, while the observed source pairs $(y^{(k)},(\tbx^{(k)})^{\top})$ are labeled as $L=0$.

{\it Step 3}. Estimate the weight and the sample selection subset.  Combine the synthetic sample and the observed source sample to train a probabilistic classifier $\hat D^{(k)}(\tilde{\bz}) \in(0,1)$. 
For instance, it can be estimated by   standard logistic regression   in low-dimensional settings or regularized logistic regression  in high-dimensional settings. 
Then from Proposition \ref{prop: weight_optimality}, with $\tilde{\bz}_i^{(k)}=(y_i^{(k)}, (\tbx_i^{(k)})^{\top})$, the estimated importance weight is given by
$$
\hat\omega_i^{(k)}=\frac{\hat D^{(k)}(\tilde{\bz}_i^{(k)})}{1-\hat D^{(k)}(\tilde{\bz}_i^{(k)})}, \qquad i\in[n_k],\ k\in[K].
$$
Furthermore, with $\hat\eta_i^{(k)}=(\tbx_i^{(k)})^\top\bigl(\bG^{(k)}\bbeta_{\rm in}^{(0)}-\btheta_{\rm in}^{(k)}\bigr),$ the estimated selected sample set is defined as
$$
\setlength\abovedisplayskip{10pt}
 \setlength\belowdisplayskip{10pt}
\widehat{\mathcal I}_k=\Bigl\{i\in[n_k]:|y_i^{(k)}-(\tbx_i^{(k)})^\top\bG^{(k)}\bbeta_{\rm in}^{(0)}|\le R,\ |\hat\eta_i^{(k)}|\le M_k\Bigr\},\qquad k\in[K].
$$

We now introduce the final estimator. For $k=0$, we set $\bG^{(0)}=\bI_p$, $\tbx_i^{(0)}=\bx_i^{(0)}$, $\widehat{\mathcal I}_0=[n_0]$, and $\hat\omega_i^{(0)}=1$ for all $i\in[n_0]$.
The Heterogeneous Importance Weighting (HIW) estimator of $\bbeta^{(0)}$, denoted by $\widehat\bbeta_{{\hiw}}^{(0)}=(\hat\beta_{{\hiw},1}^{(0)},\ldots,\hat\beta_{{\hiw},p}^{(0)})^\top$, is defined as:
\begin{equation}\label{eq: ols iwtl lmmv}
\widehat\bbeta_{{\hiw}}^{(0)}=\underset{\bbeta\in\mR^p}{\argmin}\frac{1}{2N}\sum_{k=0}^K\sum_{i=1}^{n_k}\hat\omega_i^{(k)}\bigl(y_i^{(k)}-(\tbx_i^{(k)})^\top\bG^{(k)}\bbeta\bigr)^2\mathbb I(i\in\widehat{\mathcal I}_k)+ \sum_{j=1}^{p} \lambda_{j} | \beta_{j} |,
\end{equation}
where $\lambda_j$'s are tuning parameters. In particular,  we set $\lambda_j=0$ for all $j$ in low-dimensional settings to remove the penalty.

For the theoretical analysis, the initial estimators $\bbeta_{\rm in}^{(0)}$ and $\btheta_{\rm in}^{(k)}$ are required to be independent of the data used in the final transfer learning stage.
We therefore implement Algorithm \ref{alg:HIW_main} with a data-splitting strategy.
In addition, when estimating $D^{(k)}$, the synthetic sample should be independent of the observed source sample used for classifier training; this requirement is ensured by the data-splitting procedure in Algorithm~\ref{alg: CWE}.
The construction of the projection matrix $\bG^{(k)}$ in Step 2 of Algorithm \ref{alg:HIW_main} is described in Section \ref{sec: projection matrix}.

\begin{algorithm}[htbp]
\caption{The HIW Transfer Learning Algorithm}
\label{alg:HIW_main}

\textbf{Step 1 (Data splitting).}
Randomly split $[n_k]$ into two equal-sized parts, $\mathcal D_{k1}$ and $\mathcal D_{k2}$, for $k=0,1,\ldots,K$.
\\

\textbf{Step 2 (Initial estimation).}
(1) Estimate ${\bbeta}_{\rm in}^{(0)}$ using the target data in $\mathcal D_{01}$, and estimate ${\btheta}_{\rm in}^{(k)}$ using the source data in $\mathcal D_{k1}$ for $k\in[K]$; OLS and LASSO are used in the low- and high-dimensional settings, respectively.
(2) Construct the projection matrix $\bG^{(k)}$ using data in $\mathcal D_{01}$ and $\mathcal D_{k1}$.
\\

\textbf{Step 3 (Weight and selected sample set estimation).}
Apply Algorithm~\ref{alg: CWE} to obtain the weights $\hat\omega_i^{(k)}$ for $i\in\mathcal D_{k2}$. Compute
$\hat\eta_i^{(k)}= (\tbx_i^{(k)})^\top(\bG^{(k)}\bbeta_{\rm in}^{(0)}-\btheta_{\rm in}^{(k)})$
and 
$$
\widehat{\mathcal I}_{k2}=\{i\in\mathcal D_{k2}:|y_i^{(k)}-(\tbx_i^{(k)})^\top\bG^{(k)}{\bbeta}_{\rm in}^{(0)}|\le R,\ |\hat\eta_i^{(k)}|\le M_k\}.
$$
\\

\textbf{Step 4 (Transfer learning).}
Let $\tbx_i^{(0)}=\bx_i^{(0)}$, $\bG^{(0)}=\bI_p$, $\hat\omega_i^{(0)}=1$, and $\widehat{\mathcal I}_{02}=\mathcal D_{02}$.
In low-dimensional settings, one takes $\lambda_j=0$ for all $j\in[p]$.
Then solve
$$
\widehat{\bbeta}_{{\hiw},1}^{(0)}=\underset{\bbeta\in\mR^p}{\argmin}\frac{1}{2N}\sum_{k=0}^K \sum_{i\in\mathcal D_{k2}}\hat\omega_i^{(k)}\bigl(y_i^{(k)}-(\tbx_i^{(k)})^\top\bG^{(k)}\bbeta\bigr)^2\mathbb I(i\in\widehat{\mathcal I}_{k2})+\sum_{j=1}^p \lambda_j |\beta_j|.
$$
\\

\textbf{Step 5 (Cross-fitting).}
Switch the roles of $\mathcal D_{k1}$ and $\mathcal D_{k2}$, repeat Steps 2--4 to obtain $\widehat{\bbeta}_{{\hiw},2}^{(0)}$, and define the final estimator by
$\widehat{\bbeta}_{{\hiw}}^{(0)}={(\widehat{\bbeta}_{{\hiw},1}^{(0)}+\widehat{\bbeta}_{{\hiw},2}^{(0)})}/{2}.$
\end{algorithm}

\begin{algorithm}[htbp]
\caption{Classification-based Weight Estimation}
\label{alg: CWE}

\textbf{Step 1 (Data splitting).}
For each $k=1,\ldots,K$, randomly split $\mathcal D_{k1}$ into two equal-sized subsets, denoted by $\mathcal D_{k11}$ and $\mathcal D_{k12}$.
\\

\textbf{Step 2 (Synthetic data generation).}
For $i\in\mathcal D_{k11}$, draw a synthetic noise variable $\nu_i^{(k)}\sim \nu$ and generate the pseudo-response $y_i^{(k0)}=(\tbx_i^{(k)})^\top\bG^{(k)}{\bbeta}_{\rm in}^{(0)}+\nu_i^{(k)}.$
Let $\mathcal Z_1^{(k)}=\{(y_i^{(k0)},(\tbx_i^{(k)})^{\top}): i\in\mathcal D_{k11}\}$ be the synthetic dataset labeled as $L=1$.
\\

\textbf{Step 3 (Observed data).}
Let $\mathcal Z_0^{(k)}=\{(y_i^{(k)},(\tbx_i^{(k)})^{\top}): i\in\mathcal D_{k12}\}$
be the observed source dataset, with class label $L=0$.
\\

\textbf{Step 4 (Probabilistic classification).}
Train a probabilistic classifier $\hat D^{(k)}$ on the pooled dataset $\mathcal Z_0^{(k)}\cup\mathcal Z_1^{(k)}$.
We use logistic regression in low-dimensional settings and logistic regression with an $\ell_1$ penalty in high-dimensional settings.
\\

\textbf{Step 5 (Weight estimation).}
For each $k\in[K]$ and $i\in\mathcal D_{k2}$, estimate the weight by
$$
\hat\omega_i^{(k)}=\frac{\hat D^{(k)}(\tilde{\bz}_i^{(k)})}{1-\hat D^{(k)}(\tilde{\bz}_i^{(k)})}, \qquad \tilde{\bz}_i^{(k)}=(y_i^{(k)},(\tbx_i^{(k)})^\top).
$$
\end{algorithm}

It is worth noting that the convergence rate of $\hat\beta_{{\hiw},j}^{(0)}$ may vary across $j\in[p]$, because the effective sample size involved in estimating different coordinates $\beta_{j}^{(0)}$ may not be the same; see Remark~\ref{rk:2} below for an illustration. This makes it natural to study the entry-wise convergence behavior of $\widehat{\bbeta}_{{\hiw}}^{(0)}$, which is established in the next section.

\begin{remark}\label{rk:2}
Consider a simple setting where the target design matrix $\bX^{(0)}$ and the observed source design matrices $\tbX^{(k)}$ have orthogonal and normalized columns, $\bG^{(k)}=\bI_{\Theta_k,\bdot}$, $\widehat{\mathcal I}_k=[n_k]$, $\lambda_j=0$, and $\hat\omega_i^{(k)}=1$ for all $i\in[n_k]$, $k\in[K]$ and $j\in[p]$.
Let $\Xi_j=\{k\in[K]:j\in\Theta_k\}$ be the set of sources in which $X_j$ is observed.
Then
$
\hat\beta_{{\hiw},j}^{(0)}=\sum_{k\in\{0\}\cup\Xi_j}\sum_{i=1}^{n_k} y_i^{(k)}x_{ij}^{(k)},
$
where $x_{ij}^{(k)}$ is the $j$-th component of $\bx_i^{(k)}$.
Since $\Xi_j$ may vary with $j \in [p]$, different coordinates of $\widehat\bbeta_{{\hiw}}^{(0)}$ can be based on different numbers of source observations.
\end{remark}

\section{Properties of the HIW Estimator}\label{sec: theory HIW}
In this section, we study the theoretical properties of the HIW estimator $\widehat{\bbeta}_{\hiw}^{(0)}$ defined in \eqref{eq: ols iwtl lmmv} under both low- and high-dimensional settings.
This section is organized as follows. The preliminaries and regularity conditions are introduced in Section \ref{subsec: condition}, and the theoretical properties of $\widehat{\bbeta}_{\hiw}^{(0)}$ are established in Section \ref{subsec: theory}.

\subsection{Preliminaries and Regularity Conditions} \label{subsec: condition}
Let $\mathcal S_0=\{j:\beta_j^{(0)}\neq 0\}$ denote the support set of $\bbeta^{(0)}$, and let $|\mathcal S_0|=s_0$.  
In the high-dimensional setting, we assume that $s_0 \ll n_0$; in the low-dimensional setting, we set $\mathcal S_0=[p]$ and $s_0=p$.
Throughout, the number of sources $K$ is assumed to be fixed.

We next state a condition on the initial estimators.
Because the synthetic sample in Algorithm \ref{alg: CWE} is generated using $\bbeta_{\mathrm{in}}^{(0)}$ rather than $\bbeta^{(0)}$, the estimated classifier $\hat D^{(k)}$ converges to an intermediate classifier $\tilde D^{(k)}$, obtained from \eqref{eq: odds_ratio} by replacing $\bbeta^{(0)}$ with $\bbeta_{\mathrm{in}}^{(0)}$.
\begin{condition}\label{Con: sigma and beta rate}
	(a) For $k \in [K]$, the initial estimators $\bbeta^{(0)}_{\rm in}, \btheta_{\rm in}^{(k)}$ and $\hat D^{(k)}(\tilde{\bz}^{(k)})$ are consistent and have convergence rates such that
	$\|\bbeta^{(0)}_{\rm in} - \bbeta^{(0)}\|_1 = O_p(\gamma_0),
	\|\btheta_{\rm in}^{(k)} - {\btheta}^{(k)}\|_1 = O_p(\gamma_k)$ and $\sup_{\tilde{\bz}^{(k)} }\big|\hat D^{(k)}(\tilde{\bz}^{(k)})-\tilde D^{(k)}(\tilde{\bz}^{(k)})\big|=O_p(q_k).$
	(b) These initial estimators and the projection matrices $\bG^{(k)}$ are independent of the data used in the transfer learning estimator.
\end{condition}

Condition \ref{Con: sigma and beta rate} (a) is mild. Convergence rates for initial estimators built from popular baseline procedures are outlined in Remark \ref{initial}. In transfer learning, the source sample sizes $n_k$ are typically much larger than the target sample size $n_0$, so $\gamma_k$ and $q_k$ ($k\in[K]$) are generally much smaller than $\gamma_0$. Condition \ref{Con: sigma and beta rate} (b) is satisfied by the data-splitting procedure described in Section \ref{sec: transfer}.

\begin{remark}\label{initial}
	In high-dimensional settings, let $s_k$ denote the number of nonzero elements in $\btheta^{(k)}$, with $s_k\le p_k$.
	For the initial estimators $\bbeta_{\rm in}^{(0)}$ and $\btheta_{\rm in}^{(k)}$, one may take $\gamma_0=p/\sqrt{n_0}$ and $\gamma_k=p_k/\sqrt{n_k}$ when OLS is used \citep{mourtada2022exact}, and $\gamma_0=s_0\sqrt{\log p/n_0}$ and $\gamma_k=s_k\sqrt{\log p_k/n_k}$ when the LASSO is used \citep{bickel2009simultaneous}.
	For $\hat D^{(k)}(\tilde{\bz}^{(k)})$, we assume for simplicity that $\tilde{\bz}^{(k)}=(y^{(k)},(\tbx^{(k)})^\top)$ is uniformly bounded. Under a logistic model in the low-dimensional setting and a sparse logistic model with sparsity $s_{D,k}$ in the high-dimensional setting, one typically has $q_k=(p_k+1)/\sqrt{n_k}$ and $q_k=s_{D,k}\sqrt{\log(p_k+1)/n_k}$, respectively \citep{bach2010self,negahban2012unified}.

\end{remark}

For technical convenience in the proofs, for each $k\in[K]$, we define the auxiliary set
$$
   \setlength\abovedisplayskip{10pt}
 \setlength\belowdisplayskip{10pt}
\Phi_k=\{i\in[n_k]:|\eta_i^{(k)}|\le M_k\}.
$$
Compared with $\mathcal I_k$, the auxiliary set $\Phi_k$ omits the constraint $|y_i^{(k)}-(\tbx_i^{(k)})^\top\bG^{(k)}\bbeta^{(0)}|\le R$.
Since $R$ is chosen to be sufficiently large, $\Phi_k$ and $\mathcal I_k$ have comparable sizes; see Lemma~A.3 in Appendix~A. 
For $i\in[n_k]$ and $k\in[K]$, let $\bx_{\mathrm{im},i}^{(k)}=(\bG^{(k)})^\top\tbx_i^{(k)}$.
Based on the auxiliary sets $\{\Phi_k\}_{k=1}^K$, we define
$$
   \setlength\abovedisplayskip{10pt}
 \setlength\belowdisplayskip{10pt}
\bSig^{\rm sel} = (\sigma^{\rm sel}_{j,l})=\alpha_0\mE\big[\bx^{(0)}(\bx^{(0)})^\top\big]+\sum_{k=1}^K\alpha_k\mE\big[\bx_{\mathrm{im},i}^{(k)}(\bx_{\mathrm{im},i}^{(k)})^\top \mathbb{I}(i\in \Phi_k)\big],
$$
which is the post-selection counterpart of the covariance matrix $\bGamma$ defined in \eqref{eq: gamma}. 
When $R$ is sufficiently large, the target data and the selected imputed source data can be viewed approximately as arising from a mixed distribution with covariance matrix $\bSig^{\rm sel}$.

Moreover, to account for estimation error in the selected sets, we define the lower and upper comparison sets of $\Phi_k$ as
$$
   \setlength\abovedisplayskip{8pt}
 \setlength\belowdisplayskip{8pt}
\Phi_k^-=\{i\in[n_k]:|\eta_i^{(k)}|\le M_k-\delta_n\},
\qquad
\Phi_k^+=\{i\in[n_k]:|\eta_i^{(k)}|\le M_k+\delta_n\},
$$
where $\delta_n=2\mu_{\max}\max\limits_{0\le k\le K}\gamma_k$ and $\mu_{\max}=\max\limits_{1\le k\le K}\max\limits_{1\le i\le n_k}\|\tbx_{i}^{(k)}\|_{\infty}$. 
The quantity of $\mu_{\max}$ is discussed further in Remark~\ref{rem: mu max}. 
Lemma~A.5 in Appendix~A shows that $\delta_n=o_p(1)$ under mild conditions, so the comparison sets differ from $\Phi_k$ only by a vanishing margin. 
We impose the following condition.

\begin{condition}\label{Con: rsc}
		Assume that (a) $\max_{j\in \mS_0}\sigma^{\rm sel}_{j,j}$ is upper bounded and that the smallest eigenvalue of $\bSig^{\rm sel}_{\mS_0,\mS_0}$ is bounded away from zero; (b) the same condition holds when $\bSig^{\rm sel}$ is defined by replacing $\Phi_k$ with either $\Phi_k^{-}$ or $\Phi_k^{+}$.
\end{condition}

The upper bound on $\max_{j\in \mS_0}\sigma^{\rm sel}_{j,j}$ in Condition \ref{Con: rsc} is mild. 
The assumption on $\bSig^{\rm sel}_{\mS_0,\mS_0}$ prevents collinearity among the covariates of index $\mS_0$ for the selected data. 
It is a variant of the standard minimum eigenvalue condition commonly used in both low-dimensional regression where $\mS_0=[p]$ \citep{mourtada2022exact}, and high-dimensional sparse regression \citep{bickel2009simultaneous}.
Part (b) is mild because $\Phi_k^{-}$ and $\Phi_k^{+}$ have the same order of size as $\Phi_k$ with probability tending to one; see Lemma~A.6 in Appendix~A.

\begin{remark}\label{rem: mu max}
	If the covariates $\tbx^{(k)}$ are sub-Gaussian, then $\mu_{\max}=\max\limits_{1 \leq k \leq K}O_p\{\sqrt{\log (n_k p_k)}\}$; if they are bounded in $\ell_\infty$ norm, then $\mu_{\max}=O(1)$. 
	The bounded-covariate condition is commonly encountered in applications such as image data and gene-expression data \citep{shorten2019survey,vinas2022adversarial}.
\end{remark}

\subsection{Theoretical Results}\label{subsec: theory}
We now present the main theoretical results for the HIW estimator $\widehat{\bbeta}_{\hiw}^{(0)}$, starting with the low-dimensional regime and then extending to the high-dimensional case.
For clarity, we introduce the following condition for projection matrices.
	While Proposition~\ref{prop: iwtl} assumes $\bG^{(k)}=P_\bB^{(k)}$, this form can be relaxed as stated in Remark~\ref{rem: iwtl}; all subsequent theoretical results hold for any $\bG^{(k)}$ satisfying Condition~\ref{Con: gk bounded}. 
\begin{condition}\label{Con: gk bounded}
For each $k \in [K]$, (a) $\bG_{\bdot, \Theta_k}^{(k)} = \bI_{p_k},$ where $\bI_{p_k} \in \mathbb{R}^{p_k \times p_k}$ is the identity matrix, and (b) $\interleave\bG^{(k)}\interleave_1$ is bounded.
 \end{condition}

Part (a) of Condition~\ref{Con: gk bounded} keeps the observed covariates unchanged after imputation. 
Part (b) prevents excessive inflation of the imputed covariates since $\|(\tbx^{(k)})^{\top}\bG^{(k)}\|_{\infty} \lesssim \|\tbx^{(k)}\|_{\infty}$.
We then introduce the quantities needed for the convergence rates. Let $n_{\mathcal I_k}=|\mathcal I_k|$ and, for each $j\in[p]$, let $\Xi_j=\{k\in[K]:j\in\Theta_k\}$ be the sources in which $X_j$ is observed. 
Define the post-selection covariance matrix $\bSig_{\mI}^{(k)}=\mE[\tbx_i^{(k)}(\tbx_i^{(k)})^\top\mid i\in\mI_k] \in \mR^{p_k \times p_k}$.
For the $j$-th covariate, define the effective sample size
$N_j=n_0+N_j^{\rm dir}+N_j^{\rm imp},$
where
$$
 \setlength\abovedisplayskip{10pt}
 \setlength\belowdisplayskip{10pt}
  N_j^{\rm dir}=\sum\limits_{k\in \Xi_j}^{}\mE(n_{\mI_{k}})\big[\bSig_{\mI}^{(k)}\big]_{j,j},\quad
  N_j^{\rm imp}=\sum\limits_{k\notin \Xi_j}\mE(n_{\mI_{k}})(\bG_{\bdot,j}^{(k)})^\top\bSig_{\mI}^{(k)}\bG_{\bdot,j}^{(k)}.
$$
Here, $N_j^{\rm dir}$ and $N_j^{\rm imp}$ denote the contributions from directly observed and imputed source covariates, respectively.
Finally, let $\boldsymbol{\delta}^{(k)}=\boldsymbol{\theta}^{(k)}-\bG^{(k)}\bbeta^{(0)}$ and $h_k=\|\boldsymbol{\delta}^{(k)}\|_1$. The next theorem gives the low-dimensional rate, where $\lambda_j=0$ for all $j\in[p]$ and $\mS_0=[p]$.
\begin{theorem}[low-dimensional case]\label{the: ld rate}
    Assume that Conditions \ref{Con: covariance epsilon}-\ref{Con: gk bounded} hold and that $p\ll \min\limits_{0\leq k\leq K}n_k$.
      Moreover, suppose that (a) $\mu_{\max}=O(1)$ and (b) $n_k \gg n_0$ such that $\gamma_k\ll \gamma_0$ and $q_k\ll \gamma_0$ for $k\in[K]$.
Under these assumptions, the HIW estimator satisfies the following entry-wise convergence rate:
    $$
     \setlength\abovedisplayskip{10pt}
 \setlength\belowdisplayskip{10pt}
    (\hat\beta^{(0)}_{{\hiw},j}-\beta_{j}^{(0)})^2=O_p\left(\frac{1}{N_j}(1+g)\right), \; j\in [p],
 $$
where $g=\sum_{k=1}^{K}\mE(n_{\mI_k})\left[\gamma_0 \wedge \left(M_k + h_k\right)\right]^2$.
Furthermore, the HIW estimator satisfies the following global convergence rate:
  $$
   \setlength\abovedisplayskip{10pt}
 \setlength\belowdisplayskip{10pt}
  \|\widehat{\bbeta}^{(0)}_{{\hiw}}-\bbeta^{(0)}\|_2^2= O_p\left(\sum\limits_{j=1}^{p}\frac{1}{N_j}(1+g)\right).
  $$
\end{theorem}
\begin{remark}
Conditions $(a)$ and $(b)$ in Theorem \ref{the: ld rate} are mild, as discussed in Remarks \ref{rem: mu max} and \ref{initial}, and are imposed only to streamline the presentation. A more general version of Theorem \ref{the: ld rate}, which does not require these two conditions, is given as Theorem~A.2 in Appendix~A.3.
\end{remark}
\begin{remark}\label{rem: Nj}
The definition of $N_j$ can be simplified.
If the eigenvalues of $\bSig_{\mI}^{(k)}$ are uniformly bounded away from zero and infinity, then
$$
   \setlength\abovedisplayskip{8pt}
 \setlength\belowdisplayskip{8pt}
N_j^{\rm dir}\asymp \sum_{k\in \Xi_j}\mE(n_{\mI_k}),\qquad
N_j^{\rm imp}\asymp \sum_{k\notin \Xi_j}\mE(n_{\mI_k})\|\bG_{\bdot,j}^{(k)}\|_2^2.
$$
Thus $N_j^{\rm dir}$ is of the order of the selected sample size from sources observing $X_j$, whereas $N_j^{\rm imp}$ is the selected sample size from sources imputing $X_j$, scaled by $\|\bG_{\bdot,j}^{(k)}\|_2^2$. 
Condition~\ref{Con: gk bounded}, together with the nondegeneracy assumption on the columns of $\bG^{(k)}$, implies $\|\bG_{\bdot,j}^{(k)}\|_2^2\asymp 1$ and hence $N_j\asymp n_0+\sum_{k=1}^K\mE(n_{\mI_k})$. 
\end{remark}

Theorem \ref{the: ld rate} highlights three key mechanisms of the proposed HIW estimator in the low-dimensional regime.
First, the entry-wise convergence rate is governed by the coordinate-specific effective sample size $N_j$, which captures the amount of transferable information available for $X_j$.
Since the set $\Xi_j$ may vary across coordinates, $N_j$ may also be different across $j\in[p]$; see also Remark  \ref{rem: Nj}. 

Second, the factor $g$ captures the cost of estimating the weights and selected sets. 
For simplicity, we use the simple expression $N_j\asymp n_0+\sum_{k=1}^K\mE(n_{\mI_k})$ from Remark~\ref{rem: Nj}.
If the biases $h_k$ are large, then $g$ is no more than $\sum_{k=1}^K\mE(n_{\mI_k})\gamma_0^2$, where $\gamma_0^2$ is the rate of the target-only initial estimator $\bbeta_{\rm in}^{(0)}$. 
In this case, HIW retains the target-only rate and thus avoids negative transfer. 
If the biases $h_k$ are small and $M_k$ is chosen properly, then $g$ is of smaller order, allowing HIW to improve over the initial estimator.

Third, the projection matrix $\bG^{(k)}$ appears in the bias term $h_k=\|\btheta^{(k)}-\bG^{(k)}\bbeta^{(0)}\|_1$, where $\btheta^{(k)}$ is the coefficient vector of the reduced model for source $k$. 
Thus, an appropriate choice of $\bG^{(k)}$ should make this discrepancy small. This objective differs from conventional imputation, where the projection matrix is chosen to predict missing covariates from the observed source covariates, for example, $\bG^{(k)}=(\bSig^{(k)}_{\Theta_k,\Theta_k})^{-1}\bSig^{(k)}_{\Theta_k,\bdot}$.
This observation motivates our transfer-enhanced construction of $\bG^{(k)}$ in Section~\ref{sec: projection matrix}.

We next consider the high-dimensional setting, where $\bbeta^{(0)}$ is sparse and the $\ell_1$ penalty is used for regularization. 
Define the weighted Gram matrix based on the target data and the selected imputed source data as
$$
\widehat{\mT} = \frac{1}{N}\left\{ \sum_{i=1}^{n_0}\bx_i^{(0)}(\bx_i^{(0)})^\top +\sum_{k=1}^{K} \sum_{i=1}^{n_k} \bx_{\mathrm{im},i}^{(k)}(\bx_{\mathrm{im},i}^{(k)})^\top \hat{\omega}_i^{(k)} \mathbb{I}(i \in \widehat{\mathcal{I}}_k)\right\}.
$$
We impose the following condition.
\begin{condition}\label{Con: incoherence}
Assume that $\max\limits_{j \in \mS_0^c}\|(\widehat\mT_{\mS_0,\mS_0})^{-1}\widehat\mT_{\mS_0,j}\|_1 \leq \iota$ for some constant $0\le \iota<1$.
\end{condition}

Condition \ref{Con: incoherence} is a variant of the mutual incoherence condition commonly used in high-dimensional statistics \citep{wainwright2009sharp}.
It controls the dependence between covariates inside and outside $\mS_0$.
We then obtain the following high-dimensional convergence result.
For $j\in[p]$, let $\rho_j=N_j/N$, $r_j^{(1)}=\rho_j\sqrt{(1+g)\log s_0/N_j}$, and $r_j^{(2)}=[\rho_j/(1-\iota)]\sqrt{(1+g)\log (p-s_0)/N_j}$, where $g$ is defined in Theorem \ref{the: ld rate}.

\begin{theorem}[Global convergence]\label{the: hd rate lambda}
      Assume that Conditions \ref{Con: covariance epsilon}-\ref{Con: incoherence} and the additional conditions (a) and (b) in Theorem \ref{the: ld rate} hold. Let  $\lambda^{(1)} \asymp \max_{j\in\mS_0} r_j^{(1)}$ and $\lambda^{(2)} \gtrsim  \lambda^{(1)} \vee \max_{j\in\mathcal S_0^c} r_j^{(2)}$. 
Choose $\lambda_j=\lambda^{(1)}$ for $j \in \mS_0$ and $\lambda_j=\lambda^{(2)}$ for $j \in \mS_0^c$.
Define $N_{\min,\mS_0}= \min\limits_{j\in \mS_0}N_j$, $N_{\max,\mS_0} = \max\limits_{j\in \mS_0}N_j$ and $\rho_{m}=N_{\max,\mS_0}/N_{\min,\mS_0}$.
Then we have that
$$
\|\widehat{\bbeta}^{(0)}_{\hiw}-\bbeta^{(0)}\|_2^2= O_p\left(\rho_{m}^2 \frac{ s_0 \log s_0}{N_{\min,\mS_0}}(1+g)\right).
$$
Furthermore, suppose that the smallest and largest eigenvalues of $\bSig_{\mI}^{(k)}$ are bounded away from zero and infinity. Then the global rate simplifies to
$$
\|\widehat{\bbeta}^{(0)}_{\hiw}-\bbeta^{(0)}\|_2^2= O_p\left( \frac{ s_0 \log s_0}{n_0+\sum_{k=1}^{K}\mE(n_{\mI_k})}(1+g)\right).
$$
\end{theorem}

Theorem~\ref{the: hd rate lambda} shows that the global error rate depends on the smallest effective sample size $N_{\min,\mS_0}$ and the imbalance factor $\rho_m=N_{\max,\mS_0}/N_{\min,\mS_0}$. 
Furthermore, under mild conditions, the rate can be simplified to depend on the term $n_0+\sum_{k=1}^K\mE(n_{\mI_k})$.
Then since $g=\sum_{k=1}^{K}\mE(n_{\mI_k})\left[\gamma_0 \wedge \left(M_k+h_k\right)\right]^2 \le \sum_{k=1}^K\mE(n_{\mI_k})\gamma_0^2$, the global rate of $\widehat{\bbeta}^{(0)}_{\hiw}$  is no slower than the target-only rate $\gamma_0^2$, up to a factor $s_0\log s_0$.
This suggests that HIW can control negative transfer.

We next show when HIW can improve over the target-only rate. 
Recall that $\bdelta^{(k)}=\btheta^{(k)}-\bG^{(k)}\bbeta^{(0)}$ and $h_k=\|\bdelta^{(k)}\|_1$, and let $h_{\min}=\min_{k\in[K]}h_k$.
Corollary~\ref{Cor: hd rate} provides sufficient conditions under which the bias term $g$ is sufficiently small, so that the global convergence rate of \(\widehat{\bbeta}^{(0)}_{\hiw}\) is faster than the target-only rate $\gamma_0^2$.
\begin{corollary}\label{Cor: hd rate}
  Assume that the conditions in Theorem \ref{the: hd rate lambda} hold and that $s_0=O(1)$.
Additionally, suppose that the following conditions are satisfied for $k\in[K]$: (a) $\tbx^{(k)}$ is Gaussian and $\bdelta^{(k)}$ is sparse in the $\ell_0$ sense;  (b) $h_{\min}\sum_{k=1}^{K}h_k=O(\log p/n)$; (c) $n_k\equiv n$. 
Let $M_k=h_{\min}$ for all $k\in[K]$.
Then $g=O(\log p)$, and hence,
$$
\|\widehat{\bbeta}^{(0)}_{\hiw}-\bbeta^{(0)}\|_2^2
=O_p\left(\frac{\log p}{N_{\min,\mS_0}}\right)
=o_p(\gamma_0^2).
$$
\end{corollary}
\begin{remark}
The sparsity of $\bdelta^{(k)}$ and the Gaussianity of $\tbx^{(k)}$ in part (a) can be replaced by suitable moment-type conditions.
For example, it suffices to assume that there exist positive constants $C_1 < 1$ and $C_2$ such that $P(|(\tbx^{(k)})^\top\bdelta^{(k)}|>M_k)\le C_1$ and $\|\bdelta^{(k)}\|_2\le C_2$ for each $k\in[K]$.
Similar arguments can be found in Proposition 1 of \citet{zheng2025transfer}.
\end{remark}

The key condition for a faster rate is $h_{\min}\sum_{k=1}^K h_k=O(\log p/n)$ in part (b).
This condition does not require all sources to be close to the target. For example, when $K=2$ and $h_1\le h_2$, it becomes $h_1h_2=O(\log p/n)$. 
This can hold either when $h_1$ is sufficiently small or when both $h_1$ and $h_2$ are moderate but their product is small. 
  The assumption $n_k\equiv n$ in part (c) is used only to simplify notation and can be relaxed to source sample sizes of the same order.

Next, we derive entry-wise convergence rates for the penalized HIW estimator.
To achieve this, we impose an additional condition. 
Let $\blambda = (\lambda_{1}, \ldots, \lambda_{p})^{\top}$, and let $\boldsymbol{\lambda}_{\mS_0}=(\lambda_{\mS_0,1},\ldots,\lambda_{\mS_0,s_0})^\top$ denote its subvector indexed by $\mS_0$.
\begin{condition}\label{Con: cor within S0} 
There exists a positive constant $\iota_1$ such that, for every $j\in[s_0]$,
$$
 \setlength\abovedisplayskip{10pt}
 \setlength\belowdisplayskip{10pt}
\sum_{l\in[s_0]}\left|\big[(\widehat\mT_{\mS_0,\mS_0})^{-1}\big]_{j,l}\right|\lambda_{\mS_0,l}
\le \iota_1 \big[(\widehat\mT_{\mS_0,\mS_0})^{-1}\big]_{j,j}\lambda_{\mS_0,j}.
$$
\end{condition}
\begin{remark}\label{rem: cor within S0}
Condition \ref{Con: cor within S0} is satisfied under simple sufficient conditions. Suppose that the eigenvalues of $\widehat\mT_{\mS_0,\mS_0}$ are bounded away from zero and infinity. Then the entries of $(\widehat\mT_{\mS_0,\mS_0})^{-1}$ are uniformly bounded, and its diagonal entries are bounded away from zero. 
Therefore, if $s_0=O(1)$ and $\{\lambda_{\mS_0,j}:j\in[s_0]\}$ are of the same order, then Condition \ref{Con: cor within S0} holds.
\end{remark}

Condition \ref{Con: cor within S0} controls correlations among the covariates in $\mS_0$.
It ensures that the penalty contributions from the other active coordinates do not dominate the $j$-th coordinate, allowing the global bound to be refined to an entry-wise bound.

\begin{theorem}[Entrywise convergence]\label{the: hd rate}
      Assume that Conditions \ref{Con: covariance epsilon}-\ref{Con: cor within S0} and the additional conditions (a) and (b) in Theorem \ref{the: ld rate} hold.
      Set $\lambda_j \asymp r_j^{(1)}$ for $j \in \mS_0$ and $\lambda_j = \lambda^{(2)}$ for $j \in \mS_0^c$, where $\lambda^{(2)}$ is chosen as in Theorem \ref{the: hd rate lambda}.
    Let $\widehat\mS_0=\{j:\hat{\beta}^{(0)}_{\hiw,j}\neq 0\}$ denote the estimated support set.   
    Then it follows that $\widehat\mS_0 \subseteq \mS_0$ and that  
    $$
    (\hat{\beta}^{(0)}_{{\hiw},j}-{\beta}_j^{(0)})^2 = O_p\left(\frac{\log s_0}{N_j}(1+g)\right), \quad   j\in\mS_0.
    $$  
    These results imply a global convergence rate $\|\widehat{\bbeta}^{(0)}_{\hiw}-\bbeta^{(0)}\|_2^2= O_p\left(\sum_{j\in \mS_0}\frac{\log s_0}{N_j}(1+g)\right)$, which is faster than that of Theorem \ref{the: hd rate lambda}.
\end{theorem}

This implies that, if $|\beta_j^{(0)}|^2\gg (1+g)\log s_0/N_j$ for all $j\in\mS_0$, then the penalized HIW estimator selects the relevant variables with probability tending to one.
The rates match those in Theorem \ref{the: ld rate} up to an additional $\log s_0$ factor, which is standard in high-dimensional sparse regression \citep{wainwright2019high}.

Note that Theorem \ref{the: hd rate} allows the tuning parameters $\{\lambda_j\}_{j=1}^p$ to vary across coordinates, thereby accommodating the coordinate-wise effective sample sizes. 
We also provide a result for the case of a common tuning parameter; see Corollary~A.1 in Appendix~A.3. 
The conclusions are analogous to those in Theorem \ref{the: hd rate}, with $\log s_0$ in the convergence rate replaced by
$
\log s_0+\frac{\log(p-s_0)}{(1-\iota)^2}.
$
\section{Numerical studies}\label{sec: sim}
In this section, we evaluate the numerical performance of the proposed methods in both low- and high-dimensional settings.
Section \ref{sec: projection matrix} describes the construction of the projection matrix.
Section \ref{subsec: sim-setting} presents the simulation setups.
Section \ref{subsec: sim-results} reports the global and entry-wise estimation errors and compares the proposed methods with target-only estimators.

\subsection{Data-fused and target-oriented projection matrices}\label{sec: projection matrix}
We next describe the practical construction of the projection matrices.
Under blockwise missingness, constructing projection matrices that are suitable for target-domain estimation is nontrivial.
We therefore adopt the following two-step construction strategy.

\textbf{Step 1.}
As established in Proposition~\ref{prop: iwtl}, the projection matrix for the $k$-th source domain is
$P_{\boldsymbol{B}}^{(k)}= \big(\boldsymbol{B}_{\Theta_k,\Theta_k}\big)^{-1} \boldsymbol{B}_{\Theta_k,\bdot}$ where $\boldsymbol{B}$ is a positive definite working matrix.
Motivated by covariance-based missing value imputation, we construct $\boldsymbol{B}$ by aggregating covariance information from the target and source domains to address blockwise missingness.
Specifically, we define
\begin{equation}\label{eq: B DF}
\setlength\abovedisplayskip{10pt}
\setlength\belowdisplayskip{10pt}
\bB^{\rm DF}=(1-\rho_1-\rho_2)\bI_p+\rho_1\bSig^{(0)}+\rho_2\bSig^{A},
\end{equation}
where $\rho_1,\rho_2\geq0$ and $\rho_1+\rho_2<1$ are tuning parameters to be selected by cross-validation.
The identity component $\bI_p$ provides a conservative baseline and serves as ridge regularization for numerical stability. 
The matrix $\bSig^{A}=(\sigma^A_{j,l})$ pools all jointly observed target and source covariance entries to provide a stable choice:
\[
\sigma^{A}_{j,l} = \frac{n_0\sigma^{(0)}_{j,l}+\sum_{k\in\Xi_{jl}}n_k\sigma^{(k)}_{j,l}} {n_0+\sum_{k\in\Xi_{jl}}n_k}, \qquad \Xi_{jl}=\{k\in[K]:j,l\in\Theta_k\}.
\]
We define the data-fused projection matrix as $\bG_{\rm DF}^{(k)}=P_{\bB^{\rm DF}}^{(k)}$ and use $\bG^{(k)}=\bG_{\rm DF}^{(k)}$ for HIW in the numerical studies.
In implementation, $\bSig^{(0)}$ and $\bSig^{A}$ are replaced by their sample counterparts; see Appendix~B.1.
Although the projection matrix obtained in this step is preliminary, Proposition~\ref{prop: iwtl} guarantees its theoretical validity.

\textbf{Step 2.}
Building on Step 1, we further introduce a target-oriented refinement to improve statistical efficiency.
The theory in Section~\ref{sec: theory HIW} shows that the convergence rates of the HIW estimator  depend on $h_k=\|\btheta^{(k)}-\bG^{(k)}\bbeta^{(0)}\|_1$. 
Therefore, a good choice of $\bG^{(k)}$ should make $\bG^{(k)}\bbeta^{(0)}$ close to $\btheta^{(k)}$. 
Equivalently, $(\tbx_i^{(k)})^\top\bG^{(k)}\bbeta^{(0)}$ should closely approximate $\mE[y_i^{(k)}\mid \tbx_i^{(k)}]=(\tbx_i^{(k)})^\top\btheta^{(k)}$. 
This motivates using the source data to calibrate the projection matrix.
To achieve this, we estimate a source-specific diagonal rescaling matrix $\diag(\btau)\in\mR^{p_k\times p_k}$ and replace $\bG^{(k)}$ with $\diag(\btau)\bG^{(k)}$.
Specifically, with an initial HIW estimator $\widehat{\bbeta}_{\hiw}^{(0)}$ based on $\bG^{(k)}$, we compute
$$
\hat{\btau}^{(k)}
=\arg\min_{\btau\in\mR^{p_k}}\left\{\frac{1}{n_k}\sum_{i=1}^{n_k}\left(y_i^{(k)}-(\tbx_i^{(k)})^\top\operatorname{diag}(\btau)\bG^{(k)}\widehat{\bbeta}_{\hiw}^{(0)}\right)^2+\lambda_\tau\|\btau\|_1\right\}.
$$
In low-dimensional settings, we set $\lambda_\tau=0$.  
The adaptive target-oriented projection matrix is then defined as $\bG_{\rm ad}^{(k)}=\operatorname{diag}(\hat{\btau}^{(k)})\bG^{(k)}$. 
Further motivation for this adaptive refinement is provided in Appendix~B.1.
In the numerical studies, we first compute the HIW estimator   $\widehat{\bbeta}_{\hiw}^{(0)}$ using the data-fused projection matrix $\bG_{\rm DF}^{(k)}$, and then use $\widehat{\bbeta}_{\hiw}^{(0)}$ and $\bG_{\rm DF}^{(k)}$ to construct $\bG_{\rm ad}^{(k)}$. 
We finally rerun HIW with $\bG_{\rm ad}^{(k)}$ and refer to the resulting estimator as AHIW. 

\subsection{Simulation settings}\label{subsec: sim-setting}
We first describe the common simulation settings in both low- and high-dimensional regimes.
Throughout, we set $K=10$ and generate data from the linear model for $k=0,1,\ldots,K$:
$$
   \setlength\abovedisplayskip{10pt}
 \setlength\belowdisplayskip{10pt}
y_i^{(k)}=(\bx_i^{(k)})^\top\bbeta^{(k)}+\epsilon_i^{(k)},\qquad \epsilon_i^{(k)}\sim \mathcal{N}(0,1),\quad \bx_i^{(k)}\sim \mathcal{N}(\boldsymbol 0_p,\bSig^{(k)}), \quad i=1,\ldots,n_k.
$$
Let $\bSig^{(k)}=(\sigma_{ij}^{(k)})$ with $\sigma_{ij}^{(k)}=(0.5-\delta_k)^{|i-j|}$, where $\delta_0=0$ and $\delta_k\sim{\rm Unif}(0.1,0.2)$ for $k\in[K]$.
Additional simulations under a sparse covariance structure are reported in Appendix~B.3.

Next, we describe the specific setups for the low-dimensional and high-dimensional settings.
 Let $\textsf{Grid}([a,b],c)$ denote the set of grid points from $a$ to $b$ with step size $c$, and let $\textsf{Sample}(A,m)$ denote an $m$-element subset sampled from $A$ without replacement.

\textbf{Low-dimensional setting.}
Let $(p,n_0,n_k)=(5,20,70)$ for $k\in[K]$ and set $\bbeta^{(0)} = 0.5\cdot \boldsymbol{1}_{p}$.
Given $K_0\in\textsf{Grid}([0,10],2)$ and $h\in \{0.6,1\}$, we generate the source coefficient vectors by setting
\begin{itemize}
  \item[($i$)] For $1\leq k\leq K_0$, set $\bbeta^{(k)} = \bbeta^{(0)} + (h /p)\cdot \boldsymbol{1}_{p}$;
  \item[($ii$)] For $K_0+1 \leq k \leq K$, set $\bbeta^{(k)} = \bbeta^{(0)} + (2.5/p) \cdot \boldsymbol{1}_{p}$.
\end{itemize}
Under this setup, the first $K_0$ sources are closer to the target than the remaining sources, with the similarity decreasing as $h$ increases.
To construct the feature spaces $\Theta_k$, we first randomly select one source index $k'\in[K]$, denoted by $k'=\textsf{Sample}([K],1)$, and set $\Theta_{k'}=[p]$. For each remaining source $k\in[K]\setminus\{k'\}$, we set
 $$
 \setlength\abovedisplayskip{8pt}
 \setlength\belowdisplayskip{8pt}
\Theta_k=[p_k], \qquad p_k=\textsf{Sample}(\{2,3,4\},1).
$$
Thus, source $k'$ observes all covariates, whereas each other source observes only the first $p_k$ covariates. 
Recall that $X_j$ denotes the $j$-th covariate. 
Under this setup, $X_1$ is observed in all sources, whereas $X_5$ is observed only in source $k^{\prime}$. Consequently, the sample size for $X_1$ is $n_0+\sum_{k=1}^K n_k=720$, while that for $X_5$ is $n_0+n_{k^{\prime}}=90$.

\textbf{High-dimensional setting.}
Let $(p,n_0,n_k)=(300,150,300)$ for $k\in[K]$ and set $\bbeta^{(0)} = (0.5 \cdot \boldsymbol{1}_{s_0}^\top, \boldsymbol{0}^\top_{p-s_0})^{\top}$ with $s_0=10$.
Given $K_0\in\textsf{Grid}([0,K],2)$ and $h \in \{0.5, 1.5\}$, the source coefficient vectors are defined as follows:
\begin{itemize}
  \item[(i)] For $1\leq k\leq K_0$, set $\beta^{(k)}_{j} = \beta^{(0)}_{j} + h/s_0$ for $j \in \textsf{Sample}(\{s_0+1,\ldots,p\},20)$, and set $\beta^{(k)}_{j} = \beta^{(0)}_{j}$ otherwise.
  \item[(ii)] For $K_0+1 \leq k \leq K$, set $\beta^{(k)}_{j} = \beta^{(0)}_{j} + 15/s_0$ for $j \in [s_0] \cup \textsf{Sample}(\{s_0+1,\ldots,p\},20)$, and set $\beta^{(k)}_{j} = \beta^{(0)}_{j}$ otherwise.
\end{itemize}
The first $K_0$ sources are closer to the target, with similarity decreasing as $h$ increases.
For each $k\in[K]$, independently draw $a_k$ from $\{4,5,6,7,8\}$ and $u_k$ from $\textsf{Grid}([30,70],1)$, and define
$$
\Theta_k=
\begin{cases}
[s_0]\cup \textsf{Sample}\left(\{s_0+1,\ldots,p\},u_k\right), & k\in \textsf{Sample}([K],4),\\
[a_k]\cup \textsf{Sample}\left(\{s_0+1,\ldots,p\},u_k\right), & \text{otherwise}.
\end{cases}
$$
Under this construction, $X_1$ is observed in all sources, while $X_{10}$ is observed in only four sources, so the sample size of $X_1$ ($n_0+10\cdot n_k=3150$) is larger than that of $X_{10}$ ($n_0+4 \cdot n_k=1350$).

\subsection{Simulation results}\label{subsec: sim-results}
We compare HIW and AHIW with target-only estimators, namely target-OLS in the low-dimensional setting and target-LASSO in the high-dimensional setting.
A simple pooling method is omitted from the main text because it performs poorly in most settings; the corresponding results are reported in Appendix~B.2.

For HIW and AHIW, we set $M_k\equiv M$ and choose the tuning parameters $R$ and $M$ by cross-validation over the grids $\textsf{Grid}([0,6],2)$ and $\textsf{Grid}([0,1],0.2)$, respectively. The parameters $(\rho_1,\rho_2)$ are selected over $\textsf{Grid}([0.05,0.95],0.1)$ subject to the constraint in \eqref{eq: B DF}. For synthetic data generation in Algorithm \ref{alg: CWE}, we take $\nu\sim \mathcal{N}(0,1)$.

The target task is to estimate $\bbeta^{(0)}$.
Let $\widehat{\bbeta}^{(0)}=(\widehat\beta_1^{(0)},\ldots,\widehat\beta_p^{(0)})^\top$ denote the estimator obtained by a given method.
We report the global $\ell_2$ error $\|\widehat{\bbeta}^{(0)}-\bbeta^{(0)}\|_2$ and the entry-wise relative absolute error ${\rm RAE}=|\widehat\beta_j^{(0)}-\beta_j^{(0)}|/|\beta_j^{(0)}|$.
All results are averaged over 500 replications and are presented separately for the low- and high-dimensional settings.

\begin{figure}[h]
  \centering
  \includegraphics[width=0.9\textwidth]{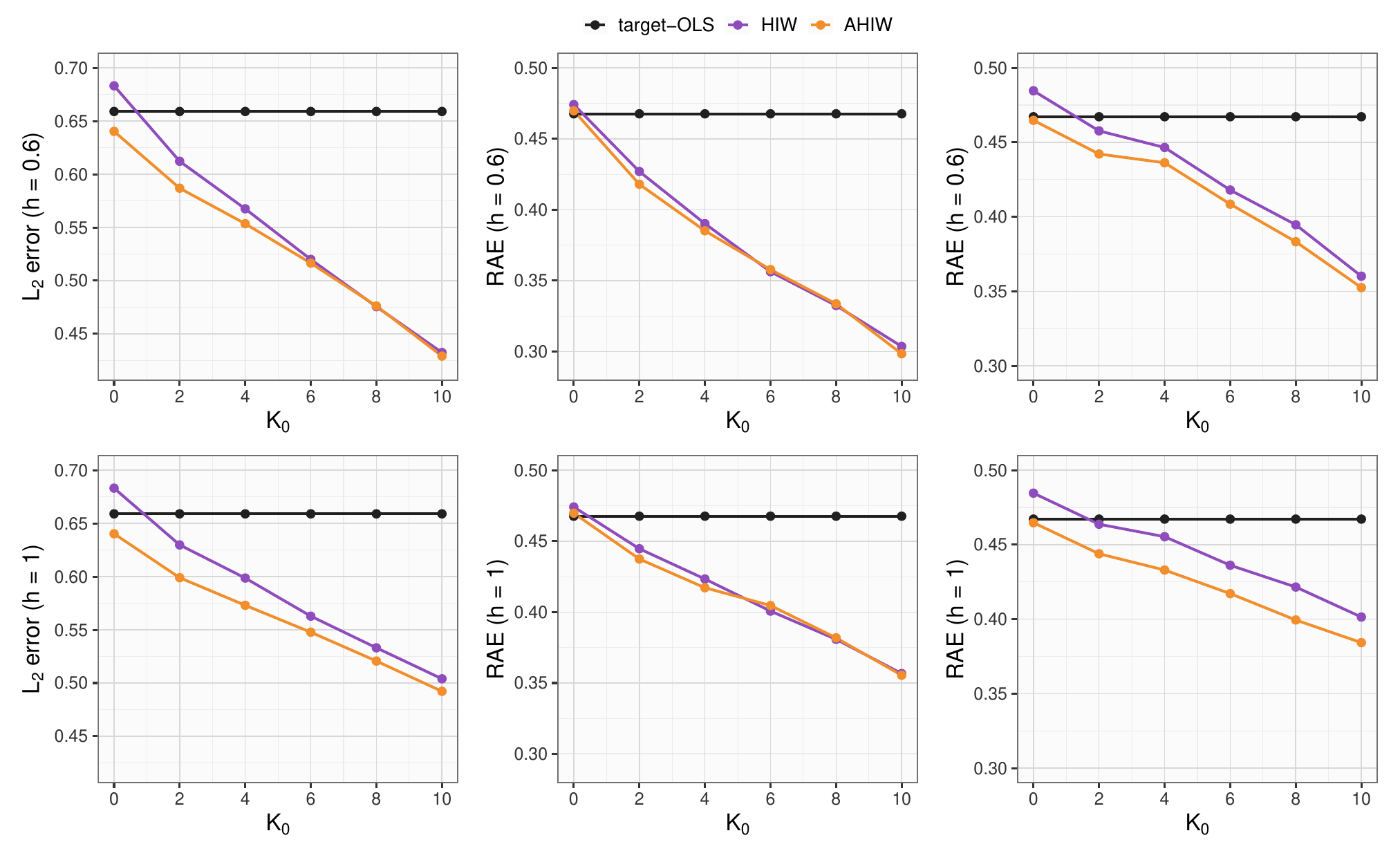}
  \caption{Numerical results in the low-dimensional setting. The first column reports the global $\ell_2$ error, while the second and third columns report the entry-wise errors (RAE) for ${\beta}^{(0)}_1$ and ${\beta}^{(0)}_5$, respectively.}
  \label{fig:ld}
\end{figure}

\textbf{Low-dimensional results.}
Figure \ref{fig:ld} summarizes the low-dimensional results: the first column reports the global estimation errors, and the second and third columns report the entry-wise errors for ${\beta}^{(0)}_1$ and ${\beta}^{(0)}_5$, respectively.
We make the following observations.

(1) \textit{Global estimation errors.}
Both HIW and AHIW improve upon target-OLS across the settings considered, indicating that the proposed transfer procedure effectively leverages source information.
Moreover, AHIW often yields smaller estimation errors than HIW in the settings considered, suggesting that adaptive refinement of the projection matrix can further improve estimation accuracy.

As $K_0$ increases, the errors of HIW and AHIW decrease because more sources are close to the target.
Similarly, smaller values of $h$ lead to smaller errors, since the informative sources are then more similar to the target.
In contrast, the performance of OLS remains essentially unchanged across different values of $K_0$ and $h$, as it uses only the target data.

(2) \textit{Entry-wise estimation errors}. The second and third columns of Figure \ref{fig:ld} report the entry-wise errors for ${\beta}_1^{(0)}$ and ${\beta}_5^{(0)}$, respectively. HIW and AHIW again outperform target-OLS. 
In most cases, AHIW performs at least comparably to HIW and often yields smaller errors.
Moreover, since the sample size of $X_1$ is larger than that of $X_5$, the error for ${\beta}_1^{(0)}$ is smaller than that for ${\beta}_5^{(0)}$, which is evident from a row-wise comparison of the second and third columns.   
In addition, AHIW, which exploits adaptive refinement of the projection matrix, exhibits greater advantages over HIW for $\beta_5^{(0)}$ relative to $\beta_1^{(0)}$.

\begin{figure}[h]
  \centering
  \includegraphics[width=0.9\textwidth]{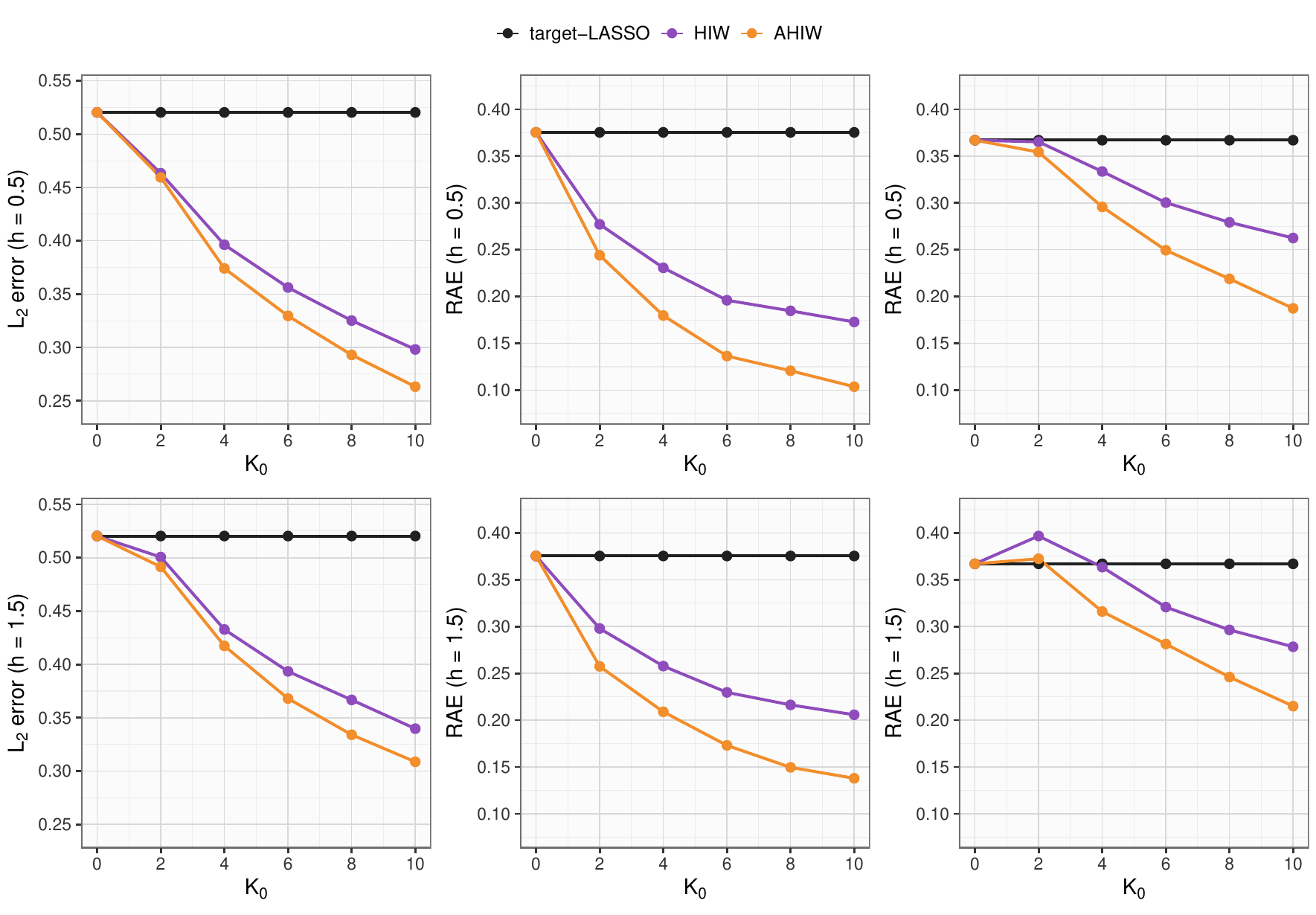}
  \caption{Numerical results in the high-dimensional setting. The first column reports the global $\ell_2$ error, while the second and third columns report the entry-wise errors (RAE) for ${\beta}^{(0)}_1$ and ${\beta}^{(0)}_{10}$, respectively.}
  \label{fig:hd}
\end{figure}

\textbf{High-dimensional results.}
Figure \ref{fig:hd} summarizes the high-dimensional results: the first column reports the global $\ell_2$ estimation errors, and the second and third columns report the entry-wise errors for ${\beta}^{(0)}_1$ and ${\beta}^{(0)}_{10}$, respectively.
We make the following observations.

(1) \textit{Global estimation errors.}
HIW and AHIW perform better than target-LASSO in most settings, suggesting that the proposed transfer procedure helps control negative transfer in the high-dimensional setting.
The errors of HIW and AHIW generally decrease as $K_0$ increases or as $h$ decreases, because more informative source data becomes available.
 Moreover, AHIW produces estimation errors comparable to or lower than those obtained by HIW.

(2) \textit{Entry-wise estimation errors}. The second and third columns of Figure \ref{fig:hd} report the entry-wise errors for ${\beta}^{(0)}_1$ and ${\beta}^{(0)}_{10}$, respectively.  
Since the sample size of $X_1$ is substantially larger than that of $X_{10}$, the estimation error of $\beta_{1}^{(0)}$ is generally smaller than that of $\beta_{10}^{(0)}$, as observed via row-wise comparison of the second and third columns.  

\section{Real data analysis}\label{sec: el}
We illustrate the proposed method using the Genotype-Tissue Expression (GTEx) dataset (\url{https://gtexportal.org/}), which contains gene expression measurements from 49 tissues and 838 human donors, yielding 1,207,976 observations for 38,187 genes. Our analysis focuses on genes associated with central nervous system neuron differentiation (GO:0021953). This pathway contains 184 genes and is documented at \url{https://www.gsea-msigdb.org/gsea/msigdb/cards/GO_CENTRAL_NERVOUS_SYSTEM_NEURON_DIFFERENTIATION}.
We take the expression level of HES5 as the response variable, since abnormal expression of this gene has been linked to developmental diseases and cancer \citep{liu2007expression}. The goal is to predict its expression level using the remaining genes in this pathway as covariates.

We restrict attention to the 13 brain tissues in GTEx. 
Among them, Hippocampus and Caudate basal ganglia are treated separately as target tissues, while the remaining 11 brain tissues are used as sources. 
The average sample size across the 13 brain tissues is about 170. 
After removing genes with missing values in the two target tissues and excluding HES5, we obtain a common target covariate set of 153 genes. 
Because many source tissues observe only a subset of these genes, the source feature spaces are naturally heterogeneous relative to the target. A summary of the numbers of observed covariates and sample sizes for all target and source tissues is provided in Appendix~C. To assess robustness under a stronger degree of feature mismatch, we additionally remove a proportion $\alpha$ of the observed covariates from each source tissue, where $\alpha \in \{0,0.3\}$.

\begin{figure}[h]
  \centering
  \includegraphics[width=0.7\textwidth]{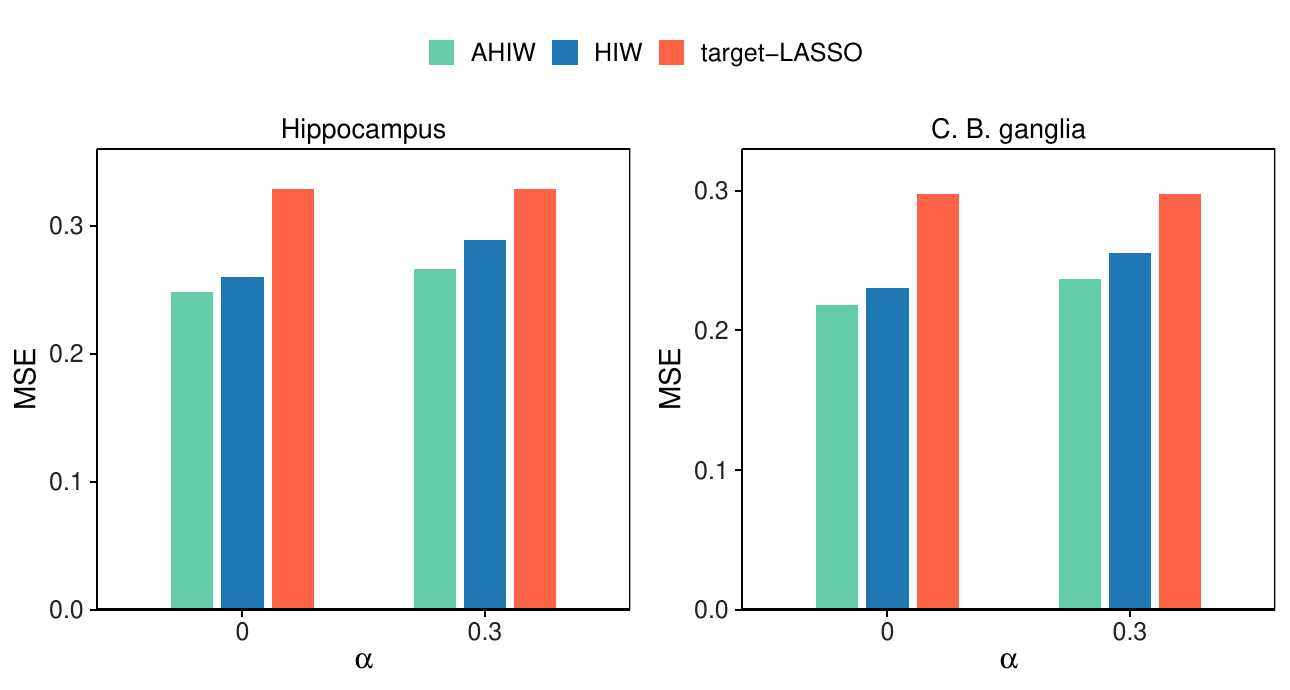}
  \caption{Mean squared prediction errors (MSE) for the two target tissues in the GTEx analysis. Here $\alpha$ denotes the proportion of additionally removed source covariates.}
  \label{fig:comparison}
\end{figure}

We compare HIW and its adaptive version AHIW with target-LASSO \citep{tibshirani1996regression}. For each target tissue, we randomly split the data into 80\% training samples and 20\% testing samples. The target-LASSO estimator is fitted using only the target training data, whereas HIW and AHIW use both target and source training data. 
For HIW and AHIW, we employ the projection matrices described in Section \ref{sec: projection matrix}. 
Prediction accuracy is evaluated by the mean squared prediction error (MSE) on the testing set, averaged over 50 replications; see Figure~\ref{fig:comparison}. The results show that both transfer estimators improve upon target-LASSO for both target tissues under $\alpha=0$ and $\alpha=0.3$. This suggests that information from related brain tissues can be effectively leveraged for prediction, even when the feature spaces are heterogeneous.
To further examine how feature mismatch affects transfer, we report the sample-selection proportions for AHIW in Appendix~C; the results show that AHIW becomes more selective when $\alpha=0.3$, indicating that stronger feature mismatch makes transfer more challenging.

\section{Discussion}\label{sec: dis}
In this paper, we address transfer learning with heterogeneous feature spaces, where source feature spaces are subsets of the target feature space. We introduce an importance weighting method for enhancing target task performance in both low- and high-dimensional linear models. Our approach combines projection-based imputation with sample selection to reduce the risk of negative transfer and is less sensitive to the choice of the projection matrix. We establish both entry-wise and global convergence rates, which can be faster than those based only on target data under suitable conditions. Future research could extend this method to generalized linear models, nonparametric regression, and cases where the target feature space is a subset of the source feature space.

\acks{This research was supported by the National Natural Science Foundation of China (Nos. 12371288 and 12131006) and the Fundamental Research Funds for the Central Universities.}

\bibliography{Bibliography-MM-MC}

\end{document}